\documentstyle[preprint,aps,epsf,graphicx]{revtex}
\begin{document}

\title{Microscopic calculations of spin polarized neutron matter 
           at finite temperature} 

\author{I. Bombaci$^1$, A. Polls$^2$, A. Ramos$^2$, A. Rios$^2$ and I. Vida\~na$^3$}

\address{$^1$Dipartimento di Fisica``E. Fermi''. Universit\`a di Pisa and INFN Sezione di Pisa, Largo Bruno
Pontecorvo
3, I-56127 Pisa, Italy}
\address{$^2$Departament d'Estructura i Constituents de la Mat\`eria. Universitat de Barcelona, Diagonal 647, E-08028 Barcelona,
Spain }
\address{$^3$Gesellschaft f\"{u}r Schwerionenforschung (GSI), Planckstrasse 1, D-64291 Darmstadt, Germany}

\date{\today}

\maketitle

\begin{abstract}

The properties of spin polarized neutron matter are studied both at zero and 
finite temperature within the framework of the Brueckner--Hartree--Fock formalism, 
using the Argonne v18 nucleon-nucleon interaction. 
The free energy, energy and entropy per particle are calculated 
for several values of the spin polarization,  densities and temperatures 
together with the magnetic susceptibility of the system.  
The results show no indication of a ferromagnetic transition at any density 
and temperature.

\vspace{0.5cm}



\end{abstract}


\vspace{1.2cm}
Since the suggestion of Pacini \cite{pa67} and Gold \cite{go68} pulsars are generally 
believed to be rapidly rotating neutron stars with strong surface magnetic fields in 
the range of $10^{12}-10^{13}$ Gauss. 
Despite the great theoretical effort of the last forty years, there is still no general 
consensus regarding the mechanism to generate such strong magnetic 
fields in a neutron star. The fields could be a fossil remnant from that of the progenitor 
star or, alternatively, they could be generated after the formation of the 
neutron star by some long-lived electric currents flowing in the highly conductive 
neutron star material. 
From the nuclear physics point of view, however, one of the most interesting and 
stimulating mechanisms which have been suggested is the possible existence of 
a phase transition to a ferromagnetic state at densities corresponding 
to the theoretically stable neutron stars and, therefore, 
of a ferromagnetic core in the liquid interior of such compact objects. 
Such a possibility has been considered since long ago by several authors within 
different theoretical approaches 
\cite{br69,ri69,cl69a,cl69b,si69,os70,pe70,pa72,ba73,ha75,ja82,ku89,ma91,be95,vi84,ku94,fa01,vi02a,vi02b,zls02,is04}, but  the results are still contradictory. 
Whereas some calculations, like for instance the ones based on Skyrme-like 
interactions  predict the transition 
to occur at densities in the range $(1-4)\rho_0$ ($\rho_0=0.16$ fm$^{-3}$), others, like recent Monte Carlo \cite{fa01} and Brueckner--Hartree--Fock  
calculations \cite{vi02a,vi02b,zls02} using modern two- and three-body realistic interactions exclude such a transition, at least up to densities around five times $\rho_0$.
This transition could have important consequences for the evolution of a protoneutron 
star, in particular for the spin correlations in the medium which do strongly affect 
the neutrino cross sections and the neutrino mean free path inside 
the star \cite{na99}. Therefore, drastically different scenarios for the evolution 
of protoneutron stars emerge depending on the existence of such a ferromagnetic 
transition.

  Most of the studies of the ferromagnetic transition in neutron and nuclear matter have 
been done at zero temperature.   However, the description of protoneutron  
stars \cite{pra97}  motivates a study of spin polarized neutron matter at temperature $T$ 
of the order of a few tens of MeV.    
Recently,  the properties of polarized neutron matter both at finite and zero temperature,  
have been investigated \cite{ri05}  using a large sample of Skyrme-like interactions.     
The results of Ref. \cite{ri05} indicate  the occurrence of a ferromagnetic phase of neutron 
matter.   However, contrary to what one would intuitively expect,  the authors of 
Ref. \cite{ri05}  have found that the critical density at which ferromagnetism takes 
place decreases with temperature. 
This unexpected result was associated to an anomalous 
behaviour of the entropy of the system  which becomes larger for the 
spin-polarized phase with respect the one for the non-polarized phase,   
above a certain density.  This was shown to be related to 
the dependence of the effective masses of neutrons with spin up and down 
on the amount of spin-polarization, 
and a new constraint on the parameters of the Skyrme force 
was derived if this anomalous behaviour is to be avoided \cite{ri05}. 

In the present work, we study the bulk and single particle properties 
of spin-polarized neutron matter at finite temperature.   
To this aim we make use of a microscopic approach based on the 
Brueckner--Hartree--Fock (BHF) approximation of the Brueckner--Bethe--Goldstone (BBG) 
expansion. Here we make use of an extension of the BBG theory  
({\it i}) to the case in which  neutron matter is arbitrarily asymmetric in the spin degree 
of freedom \cite{vi02a}   ({\it i.e.,} $\rho_\uparrow \neq \rho_\downarrow$,  where 
$\rho_\uparrow$ ($\rho_\downarrow$) is the density of neutron with spin up (down)), 
and ({\it ii}) to the case of finite temperature.       
In particular, we study the behaviour of the entropy of the system and the 
effective mass of neutrons as a function of the spin polarization parameter, 
$\Delta=(\rho_\uparrow-\rho_\downarrow)/(\rho_\uparrow+\rho_\downarrow$). 
We show that, contrary to what it is found in Ref.\ \cite{ri05}, the entropy of the 
polarized phase is lower than that of the non-polarized one, according to the idea 
that the polarized phase is more ``ordered'' than the non-polarized one. 


Our calculation starts with the construction of the neutron-neutron $G$-matrix, which 
describes in an effective way the interaction between two neutrons for each one of the spin combinations 
$\uparrow\uparrow,\uparrow\downarrow, \downarrow\uparrow$ and $\downarrow\downarrow$. This is formally obtained
by solving the well known Bethe--Goldstone equation, written schematically as

\begin{equation}
G(\omega)_{\sigma_1\sigma_2,\sigma_3\sigma_4}=
V_{\sigma_1\sigma_2,\sigma_3\sigma_4}
+\sum_{\sigma_i\sigma_j}V_{\sigma_1\sigma_2,\sigma_i\sigma_j}
\frac{Q_{\sigma_i\sigma_j}}{\omega-\varepsilon_{\sigma_i}-\varepsilon_{\sigma_j}+i\eta}
G(\omega)_{\sigma_i\sigma_j,\sigma_3\sigma_4} \ ,
\label{eq:gmat}
\end{equation}
where the first (last) two sub-indices indicate the spin projection $\sigma=\uparrow(\downarrow$) of the 
two neutrons in the initial (final) state, $V$ is the bare nucleon-nucleon interaction, 
$Q_{\sigma_i\sigma_j}$ is the Pauli operator which allows only intermediate states compatible 
with the Pauli principle, and $\omega$ is the starting energy defined as the sum of the non-relativistic 
single-particle energies, $\varepsilon_{\uparrow(\downarrow)}$, of the interacting neutrons. 

The single-particle energy of a neutron with momentum $k$ and spin projection $\sigma=\uparrow(\downarrow)$
is given by

\begin{equation}
\varepsilon_{\sigma}(k)=\frac{\hbar^2k^2}{2m}+Re[U_{\sigma}(k)] \ ,
\label{eq:spe}
\end{equation}
where the real part of the single-particle potential $U_{\sigma}(k)$ represents the averaged field ``felt'' by the neutron
due to its interaction with the other neutrons of the system. In the BHF approximation it is given by 

\begin{equation}
U_{\sigma}(k)=\sum_{\sigma'k'}n_{\sigma'}(k')\langle {\vec k}\sigma {\vec k'}\sigma' 
|G(\omega=\varepsilon_{\sigma}(k)+\varepsilon_{\sigma'}(k')|
{\vec k}\sigma {\vec k'}\sigma' \rangle _A \ ,
\label{eq:spp}
\end{equation}
where
\begin{equation}
n_{\sigma}(k) =
\left\{ \begin{array}{ll}
1,~~ \mbox{if $k \leq k^{\sigma}_{F}$} \\
0,~~ \mbox{otherwise}
\end{array} \right.
\label{eq:ocnumb}
\end{equation}
is the corresponding occupation number of a neutron with spin projection $\sigma$ and the matrix elements 
are properly anti-symmetrized. We note here that the so-called continuous prescription has been adopted for the 
single-particle potential when solving the Bethe--Goldstone equation. As shown by the authors of Refs.\ 
\cite{so98,ba00}, the contribution to the energy per particle from three-body
clusters is diminished in this 
prescription with respect to the one calculated with the gap choice for the 
single particle potential. We also note that the present calculation has been carried 
out using the Argonne v18   nucleon-nucleon potential \cite{argonne}.
The momentum dependence of the single-particle spectrum can be characterized by 
the effective mass $m^*_\sigma(k)$ defined as:
\begin{equation}
\frac{m^*_\sigma(k)}{m}=\frac{k}{m}\left(\frac{d\varepsilon_\sigma(k)}{dk}\right)^{-1} \ ,
\label{eq:mass}
\end{equation}
where $m$ is the bare neutron mass.

The total energy per particle is easily obtained once a self-consistent solution
of Eqs.\ (\ref{eq:gmat})--(\ref{eq:spp}) is achieved
\begin{equation}
\frac{E}{A} = \frac{1}{A}\sum_{\sigma k} n_{\sigma}(k)
\left(\frac{\hbar^2k^2}{2m}
+\frac{1}{2}Re[U_{\sigma}(k)]\right) \ .
\label{eq:ea}
\end{equation}

The many-body problem at finite temperature has been considered by several authors within different approaches,
such as the finite temperature Green's function method \cite{fe71}, the thermo field method \cite{he95}, or the
Bloch--De Domicis (BD) diagrammatic expansion \cite{bl58}. The latter, developed soon after the Brueckner theory,
represents the ``natural'' extension to finite temperature of the BBG expansion, to which it leads in the zero
temperature limit. Baldo and Ferreira \cite{ba99} showed that the dominant terms in the BD expansion were those
that correspond to the zero temperature of the BBG diagrams, where the temperature is introduced only through the
Fermi-Dirac distribution
\begin{equation}
f_{\sigma}(k,T)=\frac{1}{1+exp([\varepsilon_{\sigma}(k,T)-\mu_{\sigma}(T)]/T)} \ ,
\label{eq:fd}
\end{equation}
$\mu_{\sigma}(T)$ being the chemical potential of a neutron with spin projection 
$\sigma$. Therefore, at the
BHF level, finite temperature effects can be introduced in a very good approximation just replacing in the 
Bethe--Goldstone equation: (i) the zero temperature Pauli operator 
$Q_{\sigma_i\sigma_j}=(1-n_{\sigma_i})(1-n_{\sigma_j})$ by the corresponding finite temperature one
$Q_{\sigma_i\sigma_j}(T)=(1-f_{\sigma_i})(1-f_{\sigma_j})$, and (ii) the single-particle energies 
$\varepsilon_{\sigma}(k)$ by the temperature dependent ones $\varepsilon_{\sigma}(k,T)$ obtained from
Eqs.\ (\ref{eq:spe}) and (\ref{eq:spp}) when $n_{\sigma}(k)$ is replaced by $f_{\sigma}(k,T)$. These 
approximations, which are supposed to be valid in the range of densities and temperatures considered here,
correspond to the ``naive'' finite temperature Brueckner--Bethe--Goldstone (NTBBG) expansion discussed in Ref.\ \cite{ba99}.

In this case, however, the self-consistent process implies that, together with the Bethe--Goldstone equation and the 
single-particle potential, the chemical potentials of neutrons with spin up and down must be extracted at each step
of the iterative process from the normalization condition
\begin{equation}
\rho_{\sigma}=\sum_{k} f_{\sigma}(k,T) \ .
\label{eq:cp}
\end{equation}
This is an implicit equation which can be solved numerically. Note that the $G$-matrix obtained from the Bethe--Goldstone equation (\ref{eq:gmat}) and also
the single-particle potentials depend implicitly on the chemical potentials. 

Once a self-consistent solution is achieved the total free energy per particle is determined by
\begin{equation}
\frac{F}{A}=\frac{E}{A}-T\frac{S}{A} \ ,
\label{eq:free_ener}
\end{equation}
where $E/A$ is evaluated from Eq.\ (\ref{eq:ea}) replacing $n_{\sigma}(k)$ by $f_{\sigma}(k,T)$ and the
total entropy per particle, $S/A$, is calculated through the expression
\begin{eqnarray}
\frac{S}{A}=-\frac{1}{A}\sum_{\sigma k}[f_{\sigma}(k,T)\mbox{ln}(f_{\sigma}(k,T)) 
+(1-f_{\sigma}(k,T))\mbox{ln}(1-f_{\sigma}(k,T))] \ .
\label{eq:entrop}
\end{eqnarray}

From the free energy per particle, we can get the remaining macroscopic properties of the system. In our case, 
we are particularly interested in the magnetic susceptibility $\chi$, which characterizes the response of a 
system to a magnetic field and gives a measure of the energy required to produce a net spin alignment in the
direction of the field. It is given by
\begin{equation} 
\chi=\frac{\mu^2\rho}{\Big(\frac{\partial^2(F/A)}{\partial \Delta^2}\Big)_{\Delta=0}}
\end{equation}
where $\mu$ is the magnetic moment of the neutron.

\begin{figure}[htb]
\centerline{
     \includegraphics[width=\textwidth]{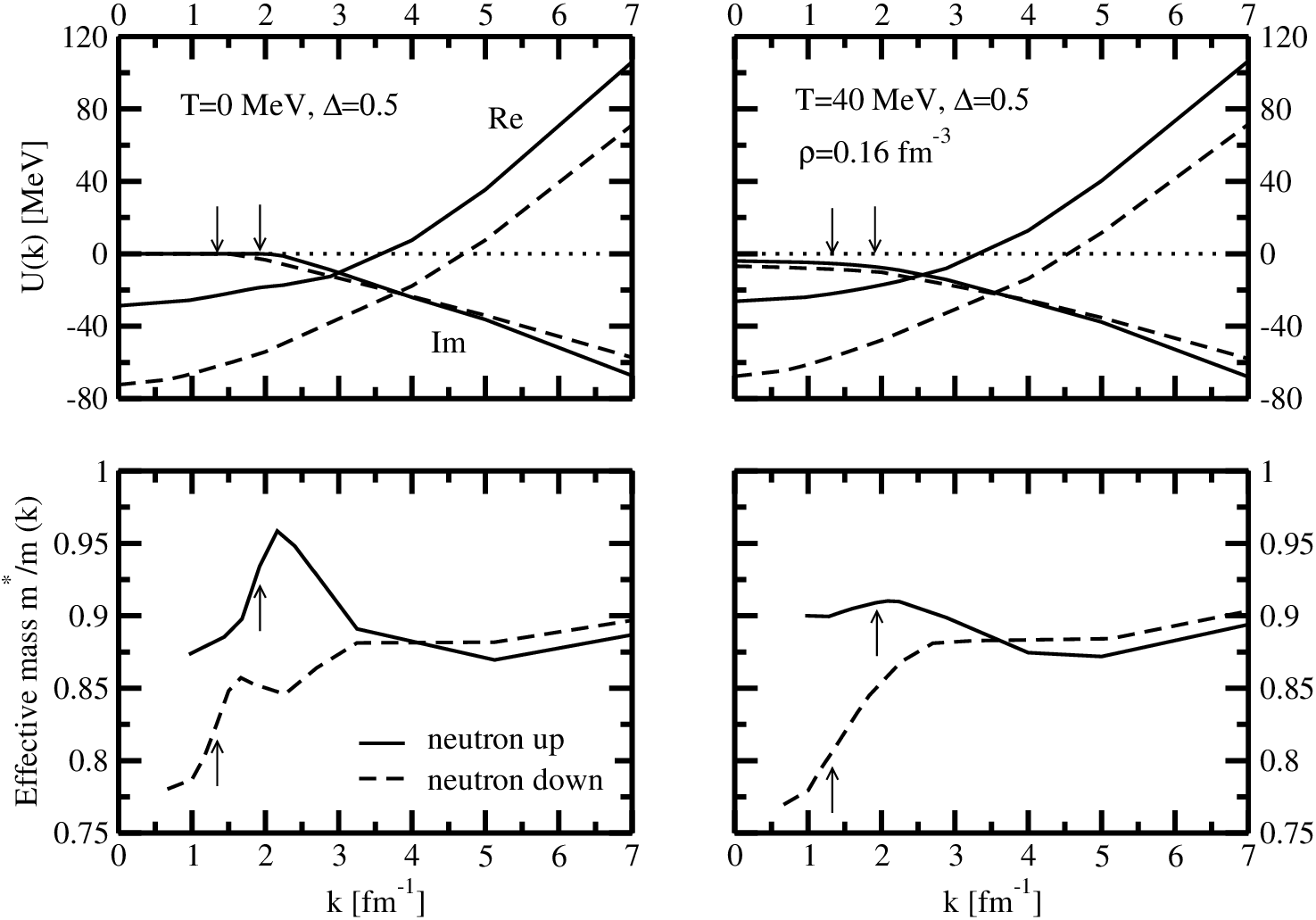}}
      \vspace{0.5cm}
      \caption{Single-particle potential (top panels) and effective mass (bottom panels) of 
neutrons with spin up (solid lines) and spin down (dashed lines) as functions  of
       the linear momentum at fixed density ($\rho=0.16$ fm$^{-3}$) and spin polarization ($\Delta=0.5$) for
       T=0 (left panels) and T=40 MeV (right panels). The arrows denote the value of the corresponding Fermi
       momenta.}
        \label{fig:fig1}
\end{figure}

The single-particle potentials of neutrons with spin up and down have been simultaneously 
and self-consistently calculated together with their effective interactions. 
The results at $\rho=0.16$ fm$^{-3}$ and spin polarization $\Delta=0.5$ are reported 
for T=0 (left panel)  and T=40 MeV (right panel) on the top panels of Fig.\ \ref{fig:fig1}. 
The neutron  single-particle potential  splits up in two different components 
when a partial spin polarization  is assumed.   In the case of Fig.\ \ref{fig:fig1}, 
the single-particle potential $Re[U_{\uparrow}(k)]$ for neutrons with spin up 
(the most abundant component)   is less attractive than the one 
for neutrons with spin down, $Re[U_{\downarrow}(k)]$.  
As demonstrated by the authors of Ref. \cite{vi02b} (see in particular their
Eqs.\ (23) and (24)), 
this splitting  ({\it i}) is  the result of a {\it phase space effect}, {\it i.e.}  to the change in the 
number of pairs which the neutron under consideration $|k,\sigma\rangle$ 
can form with the remaining neutrons 
$|k\leq k_F^{\sigma'},\sigma'=\uparrow,\downarrow \rangle $ of the system 
as neutron matter is polarized, 
and ({\it ii})  is due to the spin dependence of the neutron-neutron G-matrix in the 
spin polarized medium (see Eq.\ (\ref{eq:gmat})). Indeed, as polarization increases,
the single particle potential of a spin up neutron is built from a larger number
of up-up pairs that form  a spin triplet state ($S=1$) and, due to the Pauli principle, 
can only interact through odd angular momentum partial waves.   
Conversely, the potential of the less abundant species is built from a relatively larger 
number of up-down pairs which can interact both in the $S=0$ and $S=1$ two body 
states. Thus, the potential of the less abundant species receives also 
contributions from some important attractive channels as {\it e.g.} the $^1S_0$.

The increase of the temperature changes moderately the single-particle potentials. The real part becomes slightly less attractive, whereas the imaginary 
part increases in size as a consequence of the increase of phase space in the low momentum region. 

The momentum dependence of the corresponding effective 
masses of the two components is also shown in the bottom panels of the figure for the same values of density, spin polarization and temperatures. The general effect of temperature is
to smooth out the enhancement of the effective mass near the Fermi surface, as observed in
the work of Ref.~\cite{le86} in symmetric nuclear matter.  

\begin{figure}[htb]
\centerline{
     \includegraphics[width=\textwidth]{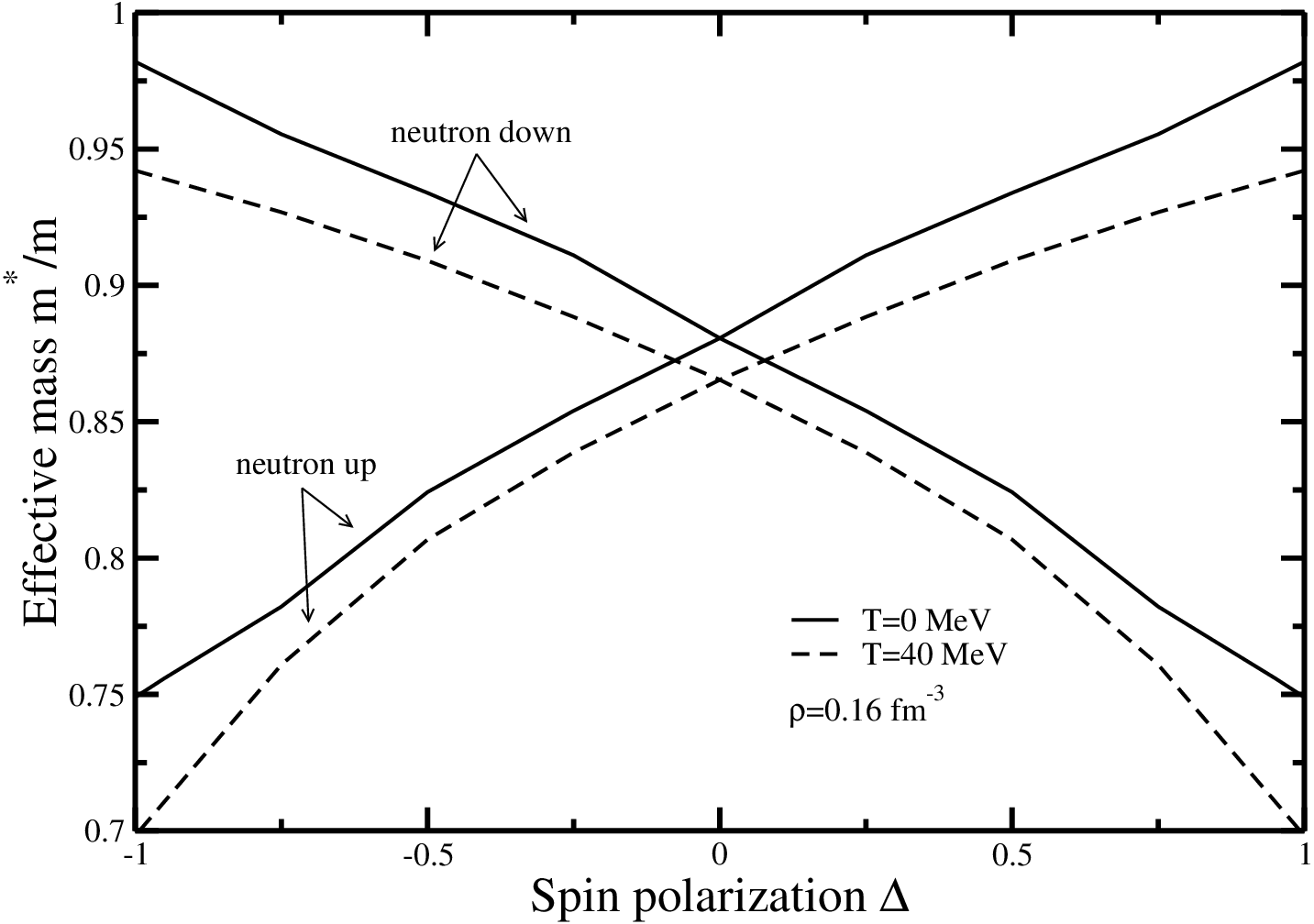}}
      \vspace{0.5cm}
      \caption{Neutron effective mass at the 
       corresponding Fermi surface of the spin up and down components as a 
       function of the spin polarization at
       $\rho=0.16$ fm$^{-3}$ for $T=0$ and $T=40$ MeV.}
        \label{fig:fig2}
\end{figure}

In Fig.\ \ref{fig:fig2} we show the effective mass $m^*_\uparrow$($m^*_\downarrow$) 
for neutron with spin up (down)  as a function of the spin polarization $\Delta$, 
for  fixed density ($\rho=0.16$ fm$^{-3}$)  and temperature  (T=0 and T=40 MeV). 
The effective mass is calculated using  Eq.\ (\ref{eq:mass}) taken for each component 
at the corresponding Fermi momentum.    
Obviously, for $\Delta=0$ the effective mass of the two components coincides.  
Once some amount of polarization is
considered, 
the values of the effective masses  split in two,  the effective mass of the most abundant 
component being larger than the one of the less abundant.  
As can be seen the effective masses show an almost linear and symmetric variation with 
respect to their  common value at spin polarization $\Delta=0$, both at T=0 and T=40 MeV. 
Deviations from this behaviour are only found at the higher polarization values.
This behaviour of $m^*_\sigma$ is a direct consequence of the scissors-like dependence 
of the single particle potential $Re[U_\sigma]$ as a function of the spin polarization 
parameter $\Delta$  (see Fig.\ 2 of Ref.\ \cite{vi02b}).   
A similar qualitative behaviour for the nucleon effective mass, as a function of the 
isospin asymmetry parameter, $\beta = (\rho_n - \rho_p)/\rho$, has been found in 
isospin asymmetric nuclear matter \cite{bo91,bo99,li04} (see in particular Eq. (94) 
in Ref. \cite{bo99}).  

\begin{figure}[htb]
\centerline{
     \includegraphics[width=\textwidth]{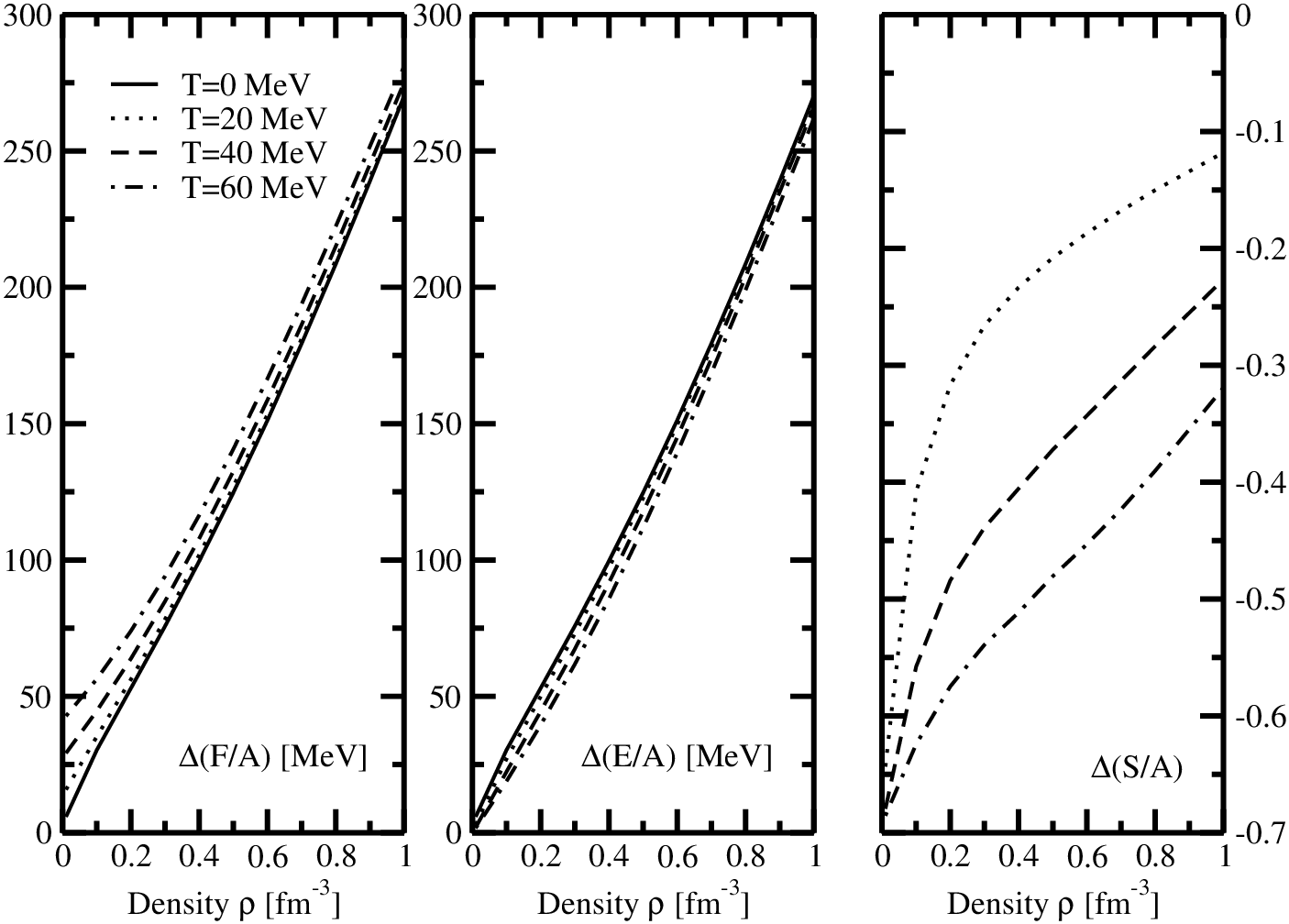}}
      \vspace{0.5cm}
      \caption{Differences of the free energy (left panel), energy (central panel)
       and entropy per particle (right panel) between fully polarized and non-polarized neutron matter as a function of density
       for several temperatures.}  
        \label{fig:fig3}
\end{figure}

The differences of the free energy (F/A), energy (E/A) and entropy (S/A) per particle between
the totally polarized and the non-polarized phases are reported in the left, central and right 
panels of Fig.\ \ref{fig:fig3} as a function of the density for several temperatures. The differences in the three quantities increase with density and increase (decrease) with
temperature in the case of the free energy (energy and entropy). 
Contrary to the results of Ref.  \cite{ri05} with the Skyrme interaction,  these 
differences are always positive for the F/A and E/A.  This is an indication that the 
non-polarized phase is energetically preferred  in the range of densities explored. 
Therefore, we can conclude that a phase transition to a ferromagnetic state is not to 
be expected from our microscopic calculation. 
If such a transition would exist, the difference in the free energy would become zero 
at some density, indicating that the ground state of the system would be ferromagnetic 
from that density on. In addition, the difference in the entropy is always negative 
indicating, as one intuitively expects, that the totally polarized phase is more ``ordered'' 
than the non-polarized one. 

\begin{figure}[htb]
\centerline{
     \includegraphics[width=\textwidth]{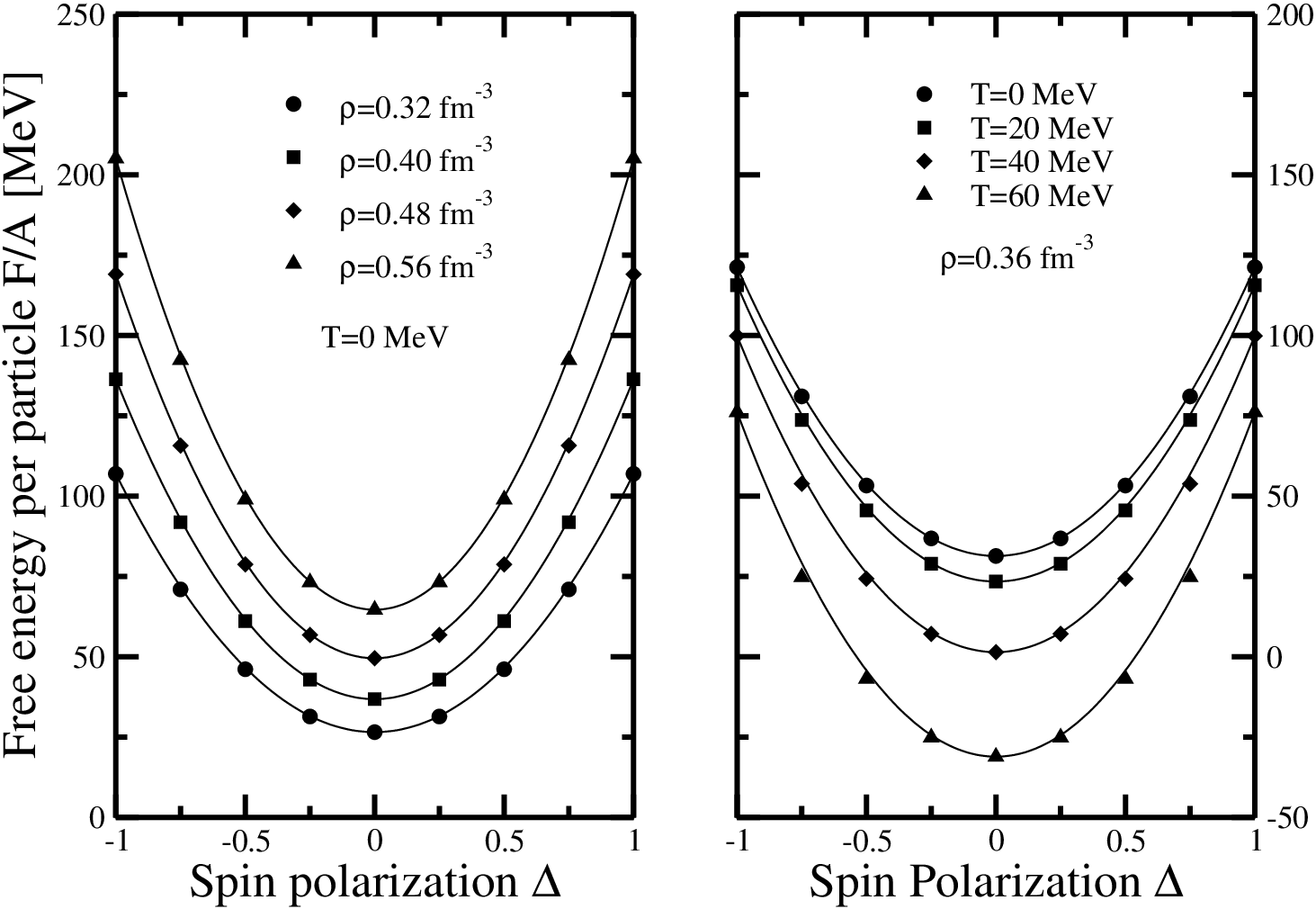}}
      \vspace{0.5cm}
      \caption{Left panel: free energy per particle at zero temperature as a
       function of the spin polarization for several densities. Right panel: free 
energy per particle at a
       fixed density $\rho=0.36$ fm$^{-3}$ as a function of the spin polarization for several temperatures.
       Circles, squares, diamonds and triangles show our BHF results, whereas the solid lines correspond to the parabolic
       approximation defined in Eq.\ (\ref{eq:para}).}
        \label{fig:fig4}
\end{figure}

In Fig.\ \ref{fig:fig4} we show the behaviour of the free energy F/A per particle as a function of the spin polarization for several densities (left panel)
and temperatures (right panel). Circles, squares, diamonds and triangles
correspond to our BHF results, whereas the solid lines correspond to the parabolic approximation
discussed below. As we expected from our previous calculations at zero temperature \cite{vi02a} and \cite{vi02b}, F/A is symmetric 
in $\Delta$ and it shows a minimum at $\Delta=0$ for all the densities and temperatures considered. This is again an indication that the ground state 
of neutron matter is paramagnetic, in opposition to what it is found in Ref. 
\cite{ri05}  
for Skyrme-like interactions where, as a consequence of the  
anomalous behaviour of the entropy, the minimum of F/A is situated at $0 < \Delta < 1$ and moves to higher polarizations when the temperature
increases. It is also interesting to note that the dependence of F/A on the spin polarization is ``up to a very good approximation'' parabolic. One can try to 
characterize that dependence in the following simple analytic form:
\begin{equation}
\frac{F}{A}(\rho, \Delta,T)=\frac{F}{A}(\rho, 0,T)+a(\rho,T)\Delta^2  
\label{eq:para}
\end{equation}
where, assuming the quadratic dependence to be valid up to $|\Delta|=1$ as our results indicate, the value of $a(\rho,T)$ can be easily obtained
for each density and temperature as the difference between the total free
energies per particle of totally polarized and  non-polarized neutron 
matter
\begin{equation}
a(\rho,T)=\frac{F}{A}(\rho, \pm1,T)-\frac{F}{A}(\rho,0,T) \ .
\label{eq:slope}
\end{equation}

The magnetic susceptibility can be evaluated then in a very simple way if the parabolic dependence of Eq.\ (\ref{eq:para}) is assumed, giving
\begin{equation}
\chi (\rho,T) = \frac{\mu^2\rho}{2a(\rho,T)} \ .
\end{equation}

\begin{figure}[htb]
\centerline{
     \includegraphics[width=\textwidth]{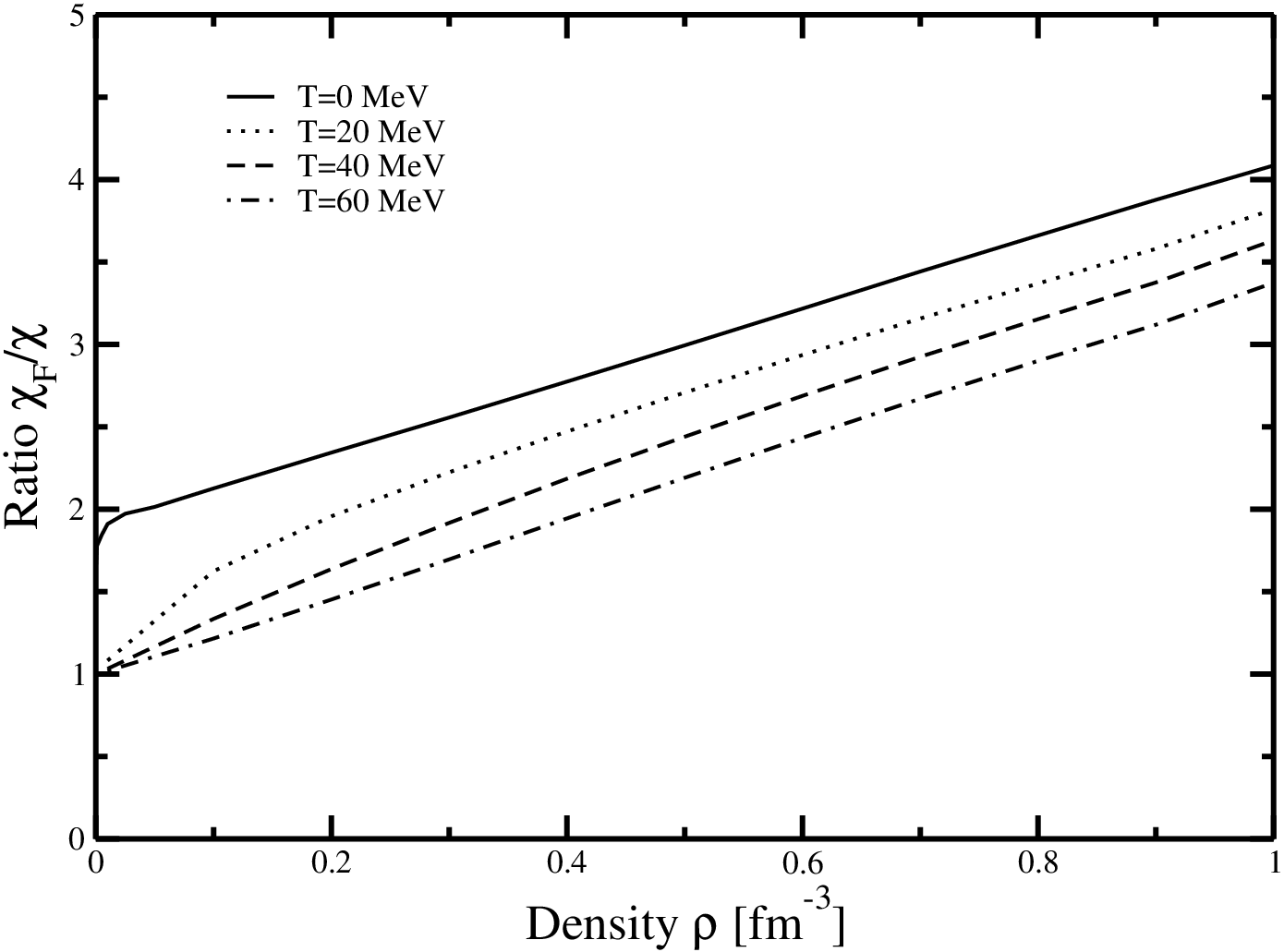}}
      \vspace{0.5cm}
      \caption{Ratio between the magnetic susceptibility of the free Fermi gas and the corresponding magnetic 
       susceptibility of interacting neutron matter as a function of density for several temperatures.}
        \label{fig:fig5}
\end{figure}

The ratio $\chi_F/\chi$, where $\chi_F$ is the magnetic susceptibility of the free Fermi gas, is shown in Fig.\ \ref{fig:fig5} as a function of density 
for several temperatures. Starting from 1, the ratio increases as the density increases 
at any temperature and no signal of a change of such a trend is expected at higher densities,  
contrary to the results of Ref. \cite{ri05} in the case of the  Skyrme-like interactions.  
This is again an indication that a ferromagnetic transition, whose onset would be signaled 
by the density at which this ratio becomes zero, is not seen and not expected at larger 
densities either. 

\begin{figure}[htb]
\centerline{
     \includegraphics[width=\textwidth]{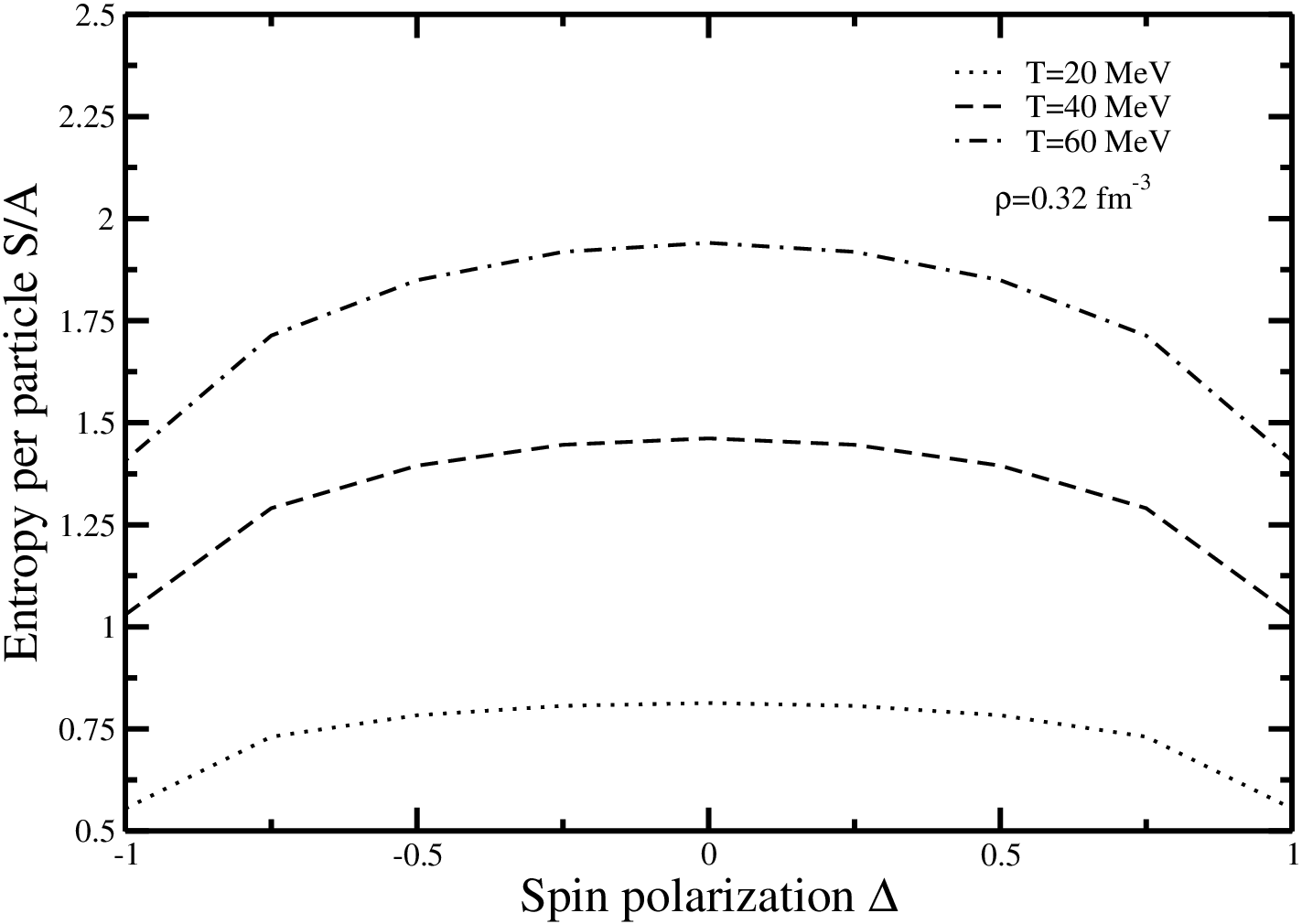}}
      \vspace{0.5cm}
      \caption{Entropy per particle as a function of the spin polarization at $\rho=0.32$ fm$^{-3}$ for
       several temperatures.}
        \label{fig:fig6}
\end{figure}

Finally, the behaviour of the entropy per particle S/A as a function of the spin polarization 
at a fixed density $\rho=0.32$ fm$^{-3}$ for several temperatures is shown in Fig.\ \ref{fig:fig6}.
The entropy, as the free energy, is also symmetric and almost 
parabolic in $\Delta$. Its  maximum is placed at $\Delta=0$ for all the densities and 
temperatures considered, as one naively expects,   contrary to the findings of Ref. \cite{ri05}. 
In this reference,  it was shown that for a pure parabolic single particle spectrum, as it is the 
case for the Skyrme interaction, imposing the entropy of the polarized phase to 
be smaller than the unpolarized one for a given density and temperature, 
is equivalent to requiring the ratio of the neutron effective masses in the 
fully polarized and  
unpolarized phases to be smaller than $2^{2/3}$.   In the BHF approach, the momentum  
and temperature dependence of the effective mass prevents from deriving a similar 
rigorous condition. However, thinking in terms of a value of the effective mass that would 
characterize the single particle spectrum in average, or considering  just the effective mass at the 
Fermi surface, which is the most relevant for the calculation of the entropy at small 
temperatures, we can then explore if the BHF calculations respect the condition 
derived in \cite{ri05}. In fact, in the case of $\rho=0.16$ fm$^{-3}$ and $T=40$ 
MeV we find (see Fig.\ \ref{fig:fig2}) 
$m^*_\uparrow(\Delta=1)/m^*_{\uparrow(\downarrow)}(\Delta=0)=1.09$, which is 
smaller than the limit established in Ref. \cite{ri05}. This is true for all the densities and temperatures explored in this work and therefore 
the entropy of the polarized phase is always smaller than that for the  unpolarized one.


In summary, 
we have studied the properties of spin polarized neutron matter both at zero and finite temperature within the
framework of the Brueckner--Hartree--Fock formalism. We have determined the single-particle potentials and the effective mass 
of neutrons with spin up and down for arbitrary values of the density, temperature and spin polarization. We have found that 
the spin up and spin down effective masses show an almost linear and symmetric variation with respect to their values at spin polarization $\Delta=0$.

We have determined the differences of the free energy (F/A), energy (E/A) and entropy 
(S/A) per particle between the totally polarized and non-polarized phases. We have found that, 
in contrast to the results of a similar study with the Skyrme interaction \cite{ri05}, 
these differences are always positive for the F/A and E/A which is an indication that the 
non-polarized phase is energetically favorable, from which we can conclude that a phase 
transition to a ferromagnetic state is not to be expected. In addition, contrary 
to the results  with the Skyrme interaction,  we have found that the difference in the entropy is 
always negative according to the idea that the totally polarized 
phase is more ``ordered'' than the non-polarized one.

Finally, we have seen that both the free energy and the entropy per particle are not 
only symmetric on the spin polarization  but also parabolic in a very good approximation 
up to $|\Delta|=1$. This finding supports the calculation of the magnetic susceptibility
by using only the free energies of the fully polarized and non-polarized phases.


\vspace{0.5cm}
This work is partially supported by DGICYT (Spain)
project BFM2002-01868 and by the Generalitat de Catalunya project
2001SGR00064. This research is part of the EU Integrated Infrastructure
Initiative Hadron Physics project under contract number RII3-CT-2004-506078. 
One of the authors (A. Rios) acknowledges the support from DURSI and the
European Social Funds.



\end{document}